%% file: main.tex
\documentclass[conference,a4paper]{IEEEtran}
\usepackage{xcolor}
\usepackage{balance}

\usepackage{cite}
\ifCLASSINFOpdf
  \usepackage[pdftex]{graphicx}
\else
  \usepackage[dvips]{graphicx}
\fi
\usepackage{array}
\ifCLASSOPTIONcompsoc
 \usepackage[caption=false,font=normalsize,labelfont=sf,textfont=sf]{subfig}
\else
 \usepackage[caption=false,font=footnotesize]{subfig}
\fi
\usepackage{url}
\usepackage[utf8]{inputenc}
\usepackage[T1]{fontenc}
\usepackage{amsmath,amssymb,amsfonts}
\usepackage{mathtools}
\usepackage{tikz}
\usepackage[capitalise]{cleveref}
\usepackage{siunitx}
 
\begin{document}
\title{On the Parametrization and Statistics\\of Propagation Graphs}

\author{\IEEEauthorblockN{
  Richard Pr\"uller\IEEEauthorrefmark{1}\IEEEauthorrefmark{2},   % 1st author, 1st affiliations
  Thomas Blazek\IEEEauthorrefmark{3},   % 2nd author, 2nd affiliations
  Stefan Pratschner\IEEEauthorrefmark{1}\IEEEauthorrefmark{2},    % 3rd author, 3rd affiliations
  Markus Rupp\IEEEauthorrefmark{2}      % 4th author, 4th affiliations
}                                     % ...
%\\
\IEEEauthorblockA{\IEEEauthorrefmark{1}CD Laboratory for Dependable Wireless Connectivity for the Society in Motion}
\IEEEauthorblockA{\IEEEauthorrefmark{2}Institute of Telecommunications, TU Wien, Vienna, Austria}
\IEEEauthorblockA{\IEEEauthorrefmark{3}Wireless Communications, Silicon Austria Labs GmbH, Linz, Austria}
\IEEEauthorblockA{\IEEEauthorblockA{Corresponding author: richard.prueller@tuwien.ac.at}}
% \IEEEauthorblockA{\IEEEauthorrefmark{3}% 3rd affiliations
% (Affiliation): dept. name of organization, name/acronyms of organization, City, Country, e-mail address*}
% \IEEEauthorblockA{\IEEEauthorrefmark{4}% 4th affiliations
% (Affiliation): dept. name of organization, name/acronyms of organization, City, Country,
%  e-mail address*}  
%  \IEEEauthorblockA{ \emph{*at least one e-mail address should be indicated above} }
}



\maketitle


\input{content.tex}

\end{document}

%% file: content.tex
%!TEX root = main.tex

% \documentclass[10pt]{article}

\usetikzlibrary{calc,arrows,arrows.meta,shapes,positioning}

\let\originalleft\left
\let\originalright\right
\renewcommand{\left}{\mathopen{}\mathclose\bgroup\originalleft}
\renewcommand{\right}{\aftergroup\egroup\originalright}

\crefformat{equation}{(#2#1#3)}
\crefmultiformat{equation}{#2(#1)#3}{ and~#2(#1)#3}{, ~#2(#1)#3}{, and~#2(#1)#3}
\crefrangeformat{equation}{#3(#1)#4--#5(#2)#6}

% \crefmultiformat{figure}{#2Fig.~#1#3}{ and~#2#1#3}{, ~#2#1#3}{, and~#2#1#3}
% \crefmultiformat{table}{#2Tables~(#1)#3}{ and~#2(#1)#3}{, ~#2(#1)#3}{, and~#2(#1)#3}
% \crefrangeformat{figure}{#3Fig.~#1#4--#5#2#6}
% \crefrangeformat{table}{#3Tables~#1#4--#5#2#6}
% \crefformat{figure}{#2Fig.~#1#3}
% \crefformat{table}{#2Table~#1#3}
% \crefformat{section}{(#2Section~#1#3)}
% \crefformat{subsection}{#2Subsection~#1#3}

\setlength{\abovedisplayskip}{6pt}
\setlength{\belowdisplayskip}{4pt}

% Vector
\let\oldv\v
\renewcommand{\v}[1]{\boldsymbol{#1}}

% Matrix
\renewcommand{\m}[1]{\boldsymbol{#1}}

% Transpose
\newcommand{\mt}[1]{{#1}^{\mathrm{T}}}
\newcommand{\mh}[1]{{#1}^{\mathrm{H}}}

\newcommand{\corrF}{\overset{\mathcal{F}}{\Longleftrightarrow}} % fourier correspondance
\newcommand{\Exp}[1]{\mathrm{E}\left\{#1\right\}} % expecation of random quantity
\newcommand{\Var}[1]{\mathrm{Var}\left\{#1\right\}} % variance of random quantity
\newcommand{\diag}[1]{\mathrm{diag}\left(#1\right)}
\newcommand{\norm}[1]{\left\lVert#1\right\rVert}

% Specific
\newcommand{\Hlos}{\m{H}_\mathrm{LOS}}
\newcommand{\Hnlos}{\m{H}_\mathrm{NLOS}}
\newcommand{\Nt}{{N_\mathrm{T}}}
\newcommand{\Nr}{{N_\mathrm{R}}}
\newcommand{\Ns}{{N_\mathrm{S}}}
\newcommand{\Gsv}{{\Gamma_\mathrm{SV}}}
\newcommand{\gsv}{{\gamma_\mathrm{SV}}}
\newcommand{\fmin}{{f_{\mathrm{min}}}}
\newcommand{\fmax}{{f_{\mathrm{max}}}}

\newcommand{\refParaOld}{\cref{eq:para12-d,eq:para12-t,eq:para12-r,eq:para12-b,eq:para12-ta,eq:para12-s}}
\newcommand{\refParaNew}{\cref{eq:new-d,eq:new-t,eq:new-r,eq:new-b}}

% -----------------------------------------------------------------------------------------------------
% \section{Abstract}
% -----------------------------------------------------------------------------------------------------

\begin{abstract}
  Propagation graphs (PGs) serve as a frequency-selective, spatially consistent channel model suitable for fast channel simulations in a scattering environment.
  So far, however, the parametrization of the model, and its consequences, have received little attention.
  In this contribution, we propose a new parametrization for PGs that adheres to the doubly exponentially decaying cluster structure of the Saleh-Valenzuela (SV) model.
  We show how to compute the newly proposed internal model parameters based on an approximation of the $K$-factor and the two decay rates from the SV model.
  Furthermore, via the singular values of multiple-input multiple-output (MIMO) channels, we compare the degrees of freedom (DoF) between our new and another frequently used parametrization.
  Specifically, we compare the DoF loss when the distance between antennas within the transmitter and receiver arrays or the average distance between scatterers decreases.
  Based on this comparison, it is shown that, in contrast to the typical parametrization, our newly proposed parametrization loses DoF in both scenarios, as one would expect from a spatially consistent channel model.
\end{abstract}

\vskip0.5\baselineskip

% -----------------------------------------------------------------------------------------------------
\section{Introduction}
% -----------------------------------------------------------------------------------------------------

Fully stochastic channel models are useful for various analytical considerations as well as higher level simulations because of their mathematical tractability and well known statistics.
However, so far there exists no fully stochastic model for wireless wide-band multiple-input multiple-output (MIMO) channels, which are becoming increasingly important.

One possible starting point for the development is a propagation graph (PG) based model, a spatially consistent wireless MIMO channel model based on scatterers.
First proposed in \cite{pedersen07}, PGs were later refined in \cite{pedersen12} and have been used successfully in several publications.
For example, in \cite{steinbock16,tian16,gan18} combinations of ray tracing and measurements verified PGs in various scenarios.
An extension of PGs to include polarization was introduced in \cite{adeogun19}.
While originally \cite{pedersen07} intended PGs as a linear time-invariant (LTI) indoor channel model, the measurement analysis in \cite{zhou15} and \cite{gan18} confirmed the applicability of PGs to time-variant outdoor scenarios using time-slicing.
A theoretical extension to the linear time-variant (LTV) case was proposed in \cite{stern18}.

In this work, we take a closer look at the parametrization of PGs and its impact on the resulting channel impulse response (CIR).
We especially draw a comparison with the Saleh-Valenzuela (SV) model \cite{saleh87}.
The SV model is still a suitable model for frequency selective channels in a random scattering environment \cite{meijerink14}, which is the same scenario PGs model.

We propose a new way to parametrize PGs, that is easier to handle from a stochastic point of view and complies with the doubly exponentially decaying cluster structure of the SV model.
Furthermore, we analytically calculate approximations for the internal model parameters as functions of the $K$-factor and the two decay rates from the SV model.

% -----------------------------------------------------------------------------------------------------
\section{Propagation Graphs}
% -----------------------------------------------------------------------------------------------------

The following section is a summary of PGs.
Typically, PGs model an LTI MIMO channel model that is spatially consistent and frequency selective.
The model is based on a finite amount of scatterers whose connections are described by a graph.
This description allows efficient handling of multiple scattering with up to infinitely many bounces.

For input vector $\v{x}(f)$, output vector $\v{y}(f)$ and channel transfer matrix $\m{H}(f)$, an LTI MIMO channel is given by
\begin{equation}\label{eq:channel}
  \v{y}(f) = \m{H}(f) \v{x}(f) = [\Hlos(f) + \Hnlos(f)] \, \v{x}(f)
\end{equation}
in the frequency domain, where $\Hlos(f)$ represent the line of sight (LOS) component and $\Hnlos(f)$ the non-line-of-sight (NLOS) component.
Let $\Nt$ be the number of Tx antennas, $\Nr$ the number of receive antennas and $\Ns$ the number of scatterers.
We define the following transfer matrices
\begin{itemize}
  \item $\m{D}(f) \in \mathbb{C}^{\Nr \times \Nt}$, from Tx to Rx,
  \item $\m{T}(f) \in \mathbb{C}^{\Ns \times \Nt}$, from Tx to scatterers,
  \item $\m{R}(f) \in \mathbb{C}^{\Nr \times \Ns}$, from scatterers to Rx,
  \item $\m{B}(f) \in \mathbb{C}^{\Ns \times \Ns}$, from scatterers to scatterers.
\end{itemize}
Using these four matrices, the LOS part of the channel transfer matrix of a PG is simply
\begin{equation}\label{eq:hlos}
  \Hlos(f) = \m{D}(f),
\end{equation}
and the NLOS part for infinite scattering bounces is
\begin{equation}\label{eq:hnlos1}
  \Hnlos(f) = \m{R}(f) \left(\sum_{k = 0}^{\infty} {\m{B}}^k(f)\right) \m{T}(f).
\end{equation}
If the spectral radius of $\m{B}(f)$ is smaller than one, i.e., the signal loses energy with each scattering bounce, then the Neumann series in \cref{eq:hnlos1} converges to \cite{pedersen07}
\begin{equation}\label{eq:hnlos2}
  \Hnlos(f) = \m{R}(f) {\left(\m{I} - \m{B}(f)\right)}^{-1} \m{T}(f).
\end{equation}
% In \cref{fig:pg} this general setup is visualized.

% \begin{figure}
%   \centering
%   \begin{tikzpicture}[->,>=stealth',shorten >=1pt,auto,thick,node distance=20mm]
%     \node[circle,draw,minimum size=10mm] (x) {$\v{x}(f)$};
%     \node[inner sep=0,minimum size=0,right of=x] (k) {};
%     \node[circle,draw,minimum size=10mm,right of=k] (y) {$\v{y}(f)$};
%     \node[circle,draw,minimum size=10mm,below of=k] (z) {};
%     \path[every node/.style={font=\small,fill=white,inner sep=1pt}]
%       (x) edge node {$\m{D}(f)$} (y)
%         edge node {$\m{T}(f)$} (z)
%       (z) edge [loop below] node {$\m{B}(f)$} (z)
%         edge node {$\m{R}(f)$} (y);
%   \end{tikzpicture}
%   \caption{General propagation graph setup}
%   \label{fig:pg}
% \end{figure}

An often used parametrization for the elements of the four transfer matrices $\m{D}(f)$, $\m{T}(f)$, $\m{R}(f)$ and $\m{B}(f)$ is \cite{pedersen12,zhou15,steinbock16,adeogun19}
\begin{align}
    D_{mn}(f) &= \frac{\varepsilon_{D,mn}}{4 \pi f \tau_{D,mn}} \cdot e^{-j 2 \pi \tau_{D,mn} f}, \label{eq:para12-d}\\
    T_{mn}(f) &= \frac{\varepsilon_{T,mn}}{\sqrt{4 \pi f \bar{\tau}_T}} \cdot \frac{\tau_{T,mn}^{-1}}{\sqrt{S_T}} \cdot e^{-j 2 \pi \tau_{T,mn} f + j \phi_{T,mn}}, \label{eq:para12-t}\\
    R_{mn}(f) &= \frac{\varepsilon_{R,mn}}{\sqrt{4 \pi f \bar{\tau}_R}} \cdot \frac{\tau_{R,mn}^{-1}}{\sqrt{S_R}} \cdot e^{-j 2 \pi \tau_{R,mn} f + j \phi_{R,mn}}, \label{eq:para12-r}\\[2pt]
    B_{mn}(f) &= \frac{g \, \varepsilon_{B,mn}}{\sum_{n = 1}^N \varepsilon_{B,mn}} \cdot e^{-j 2 \pi \tau_{B,mn} f + j \phi_{B,mn}}. \label{eq:para12-b}
\end{align}
Let $\m{A}$ be any of the four matrices from above, then $\varepsilon_{A,mn} \in \{0,1\}$ indicates whether the link connecting antenna/scatterer $m$ with antenna/scatterer $n$ is unobstructed or not.
Furthermore, $\tau_{A,mn}$ is the delay (distance divided by $c_0$) between antenna/scatterer $m$ and antenna/scatterer $n$, and $\phi_{A,mn} \sim \mathcal{U}[0,2 \pi)$ is a corresponding random phase shift uniformly distributed on $[0, 2 \pi)$.
Finally, $g \in \mathbb{R}_+$ is a model parameter and the two quantities $\bar{\tau}_A$ and $S_A$ are defined as
\begin{align}
  \bar{\tau}_A &= \frac{\sum_{m = 1}^M \sum_{n = 1}^N \varepsilon_{A,mn} \tau_{A,mn}}{\sum_{m = 1}^M \sum_{n = 1}^N \varepsilon_{A,mn}},\label{eq:para12-ta}\\[2pt]
  S_A &= \textstyle \sum_{m = 1}^M \sum_{n = 1}^N \varepsilon_{A,mn} \tau_{A,mn}^{-2}.\label{eq:para12-s}
\end{align}
Typically, all link indicators of $\m{D}$ are chosen to be equal, i.e., $\varepsilon_{D,mn} = \varepsilon_D \ \forall m, n$.
Furthermore, scatterers should not ``see'' themselves, i.e., $\varepsilon_{B,mm} = 0 \ \forall m$.

It should be noted that this parametrization was first introduced in \cite{pedersen12}, and all definitions are originally based on a graph with weighted edges.
For brevity and consistency with the sections below, however, we omitted this graph interpretation and introduced the link indicators $\varepsilon_{A,mn}$ instead.

% -----------------------------------------------------------------------------------------------------
\section{New Parametrization}
% -----------------------------------------------------------------------------------------------------

In this section, we introduce a new parametrization for PGs that replaces \refParaOld{} and fixes two potential problems.
The first problem is that according to \cref{eq:para12-t,eq:para12-r,eq:para12-b} \textit{every} scatterer to scatterer and scatterer to antenna link gets a random phase shift applied.
We show in \cref{sec:sim} and, in particular, \cref{fig:sv}, that this introduces too many DoF to the model, leading to inconsistent behavior in the MIMO case.
Secondly, the $\tau_{A,mn}^{-1}$ terms in \cref{eq:para12-t,eq:para12-r} potentially lead to a channel that amplifies if a scatterer is too close to an antenna.
However, the latter problem can be fixed for simulations by removing problematic scatterers once they have been sampled.

The SV model \cite{saleh87} is a well established, frequency selective channel model for indoor scenarios.
Since PGs model essentially the same scenario as the SV model we argue that it is desirable that PGs should reproduce the same doubly exponential decay as the SV model, i.e., the expected power of a received ray at delay $\tau$ is given by
\begin{equation}\label{eq:sv}
  P_{\mathrm{ray}}(\tau) = P_{\mathrm{ray}}(0) \sum_{n = 0}^{N} 10^{\frac{T_n \rho_1}{10}} 10^{\frac{(\tau - T_n) \rho_2}{10}} u(\tau - T_n),
\end{equation}
were $\rho_1$ is the cluster decay rate and $\rho_2$ is the ray decay rate, both in \si{dB\per{}s}.
Furthermore, $T_n$ is the arrival time of the $n$th cluster, and $u(\cdot)$ is the unit step function.

For many simulations and theoretical considerations, the $K$-factor, i.e., the fraction of the powers in the LOS and NLOS parts of the channel is important.
Denoting the expectation operator as $\Exp{\cdot}$ and the Frobenius norm with $\norm{\cdot}_\mathrm{F}$, we define
\begin{equation}\label{eq:k}
  K = \frac{P_{\mathrm{LOS}}}{P_{\mathrm{NLOS}}}
  = \frac{\int_\fmin^\fmax \Exp{\norm{\m{H}_{\mathrm{LOS}}(f)}_\mathrm{F}^2} \, df}{\int_\fmin^\fmax \Exp{\norm{\m{H}_{\mathrm{NLOS}}(f)}_\mathrm{F}^2} \, df}
\end{equation}
as the $K$-factor on a frequency interval $[\fmin, \fmax]$.

The remainder of this section is dedicated to introducing our new parametrization that addresses the above-mentioned issues and a description on how to compute the internal model parameters from the easier to use $\rho_1$, $\rho_2$ and $K$.

% -----------------------------------------------------------------------------------------------------
\subsection{The New Parametrization}

With all the above in mind, we now introduce our new parametrization of PGs.
We replace \refParaOld{} by
\begin{align}
  D_{mn}(f) &= \frac{\varepsilon_D}{4 \pi \tau_{D,mn} f} e^{-j 2 \pi \tau_{D,mn} f}, \label{eq:new-d} \\
  T_{mn}(f) &= \sqrt{\frac{\alpha}{f}} e^{\tau_{T,mn} \gamma} e^{-j 2 \pi \tau_{T,mn} f + j \phi_{T,n}}, \label{eq:new-t} \\
  R_{mn}(f) &= \sqrt{\frac{\alpha}{f}} e^{\tau_{R,mn} \gamma} e^{-j 2 \pi \tau_{R,mn} f + j \phi_{R,m}}, \label{eq:new-r} \\
  B_{mn}(f) &= (1 - \delta_{mn}) \beta e^{-j 2 \pi \tau_{B,mn} f}, \label{eq:new-b}
\end{align}
to compute the entries of the matrices in \cref{eq:hlos} and \cref{eq:hnlos2}.
Here, $\alpha$, $\beta$ and $\gamma$ are \emph{internal} model parameters and $\phi_{T,n}$ as well as $\phi_{R,m}$, $m,n = 1,\ldots,\Ns$, are i.i.d. random phases uniformly distributed on $[0,2\pi)$.

Before we proceed to compute $\alpha$, $\beta$ and $\gamma$ from $\rho_1$, $\rho_2$ and $K$, we take a closer look at this parametrization.
If we would insert \cref{eq:new-t,eq:new-r,eq:new-b} into \cref{eq:hnlos2}, we could pull out a factor $\alpha/f$, i.e.,
\begin{equation}
  \m{H}_{\mathrm{NLOS}}(f) = \frac{\alpha}{f} \tilde{\m{H}}_{\mathrm{NLOS}}(f),
\end{equation}
were $\tilde{\m{H}}_{\mathrm{NLOS}}(f)$ is no longer dependent on $\alpha$.
This means that a change to $\alpha$ only affects the \emph{magnitude} of $\m{H}_{\mathrm{NLOS}}(f)$ and thus, we can adjust the $K$-factor of the model using $\alpha$.

It was already shown that $\beta$ controls the rate of the exponential decay over time of the cluster power \cite{pedersen12,adeogun19}.
From $|B_{mn}(f)| = \beta = \mathrm{const.}$ it follows that the decay within a cluster (ray power decay) is solely defined by the delay dependency of the combined element magnitudes in $\m{R}(f)$ and $\m{T}(f)$.
Thus, we chose $\exp(\tau \gamma)$ in $T_{mn}(f)$ and $R_{mn}(f)$ to agree with the exponential power decay of the rays within a cluster of the SV model.
% A further advantage of this choice is that there is no danger of divergence for small $\tau_{T,mn}$ and $\tau_{R,mn}$, making simulations more stable.
It should also be mentioned here that $\beta$ and $\gamma$ are independent, but both have a strong influence on the $K$-factor, and thus $\alpha$, if a specific value of $K$ should be achieved.

Except for $\varepsilon_D$, which controls LOS visibility, all other link indicators were set to one.
This, in turn, means that we assume only one scatterer cluster and that all antennas and scatterers see each other.

Finally, the new parametrization restricts the additional random phase shifts $\phi$ to one per scatterer and Tx/Rx.
The reasoning behind this is as follows.
% From the perspective of a scatterer, an impinging/outgoing wave is characterized by five parameters; amplitude, phase, frequency, polarization, and direction of arrival/departure (DoA/DoD).
% It is usually safe to assume that the scattering process is independent of the amplitude and phase.
% Furthermore, dependencies of the additional phase shifts on the frequency and polarization are not modeled here.
% This leaves us with the DoA/DoD.
Assuming that Tx/Rx antennas are, respectively, in similar positions, it can be argued that for a given scatterer, all waves from/to the antennas have a similar direction of arrival/departure, and thus a similar phase shift.
This leaves us with one additional random phase shift per scatterer and Tx/Rx, i.e., in total $2 \Ns$.

\begin{figure}[t]
  \centering
  \includegraphics[width=0.95\columnwidth]{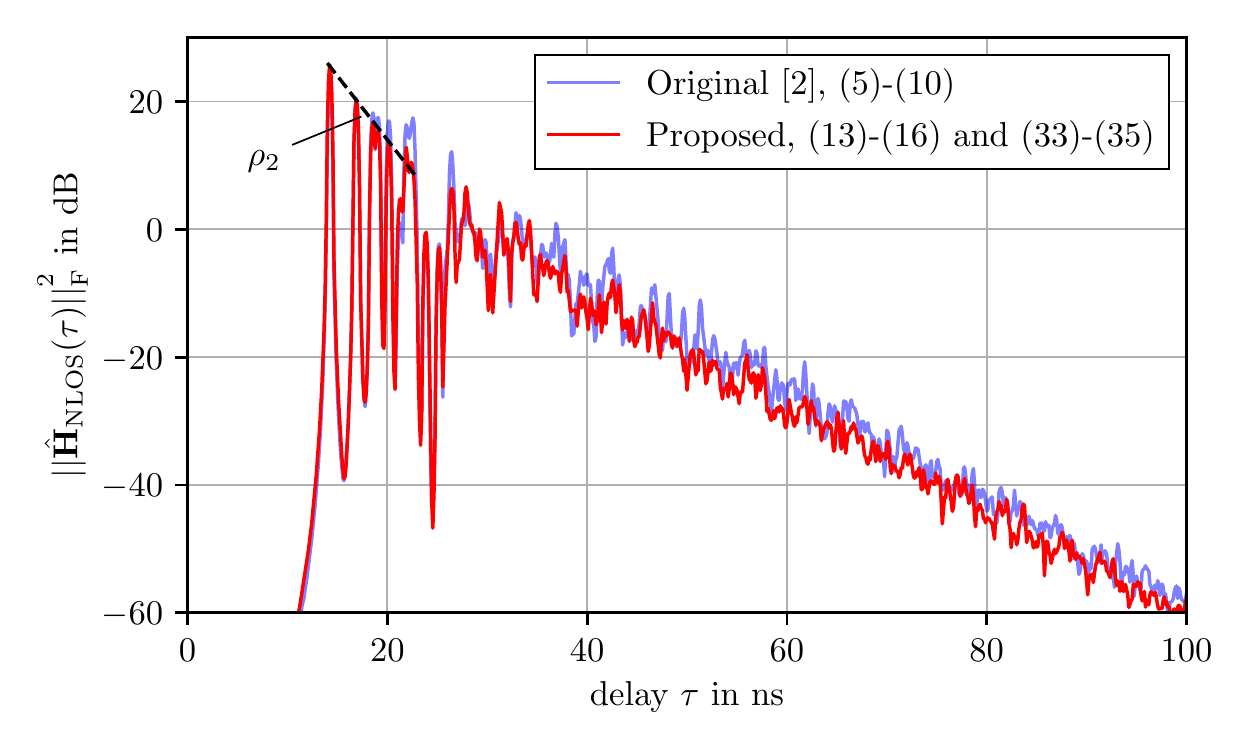}%
  \caption{%
    Comparison of a $4\times4$ MIMO realization of the NLOS channel in time domain between the original \refParaOld{} and new \refParaNew{} parametrizations using the same scatterer positions.
    The CIR $\hat{\m{H}}_\mathrm{NLOS}(\tau)$ is approximated by filtering a Hann window with $\Hnlos(f)$ and subsequent Fourier transformation as described in the appendix of \cite{pedersen12}.
  }\label{fig:realization}
\end{figure}

A qualitative comparison between the parametrizations from \refParaOld{} and \refParaNew{} is shown in \cref{fig:realization}.
Of note is that the cluster decay rates of the two models are equal, while the decay rates within the clusters differ.
This is especially visible in the first set of peaks in the CIR.
The peaks in the CIR resulting from the typical parametrization decays polynomially, and the newly proposed one exponentially.

% -----------------------------------------------------------------------------------------------------
\subsection{Derivation of the $K$-Factor}\label{ssec:deriv-k}

\begin{figure*}[t]
  \centering
  \subfloat[]{
      \includegraphics[width=0.95\columnwidth]{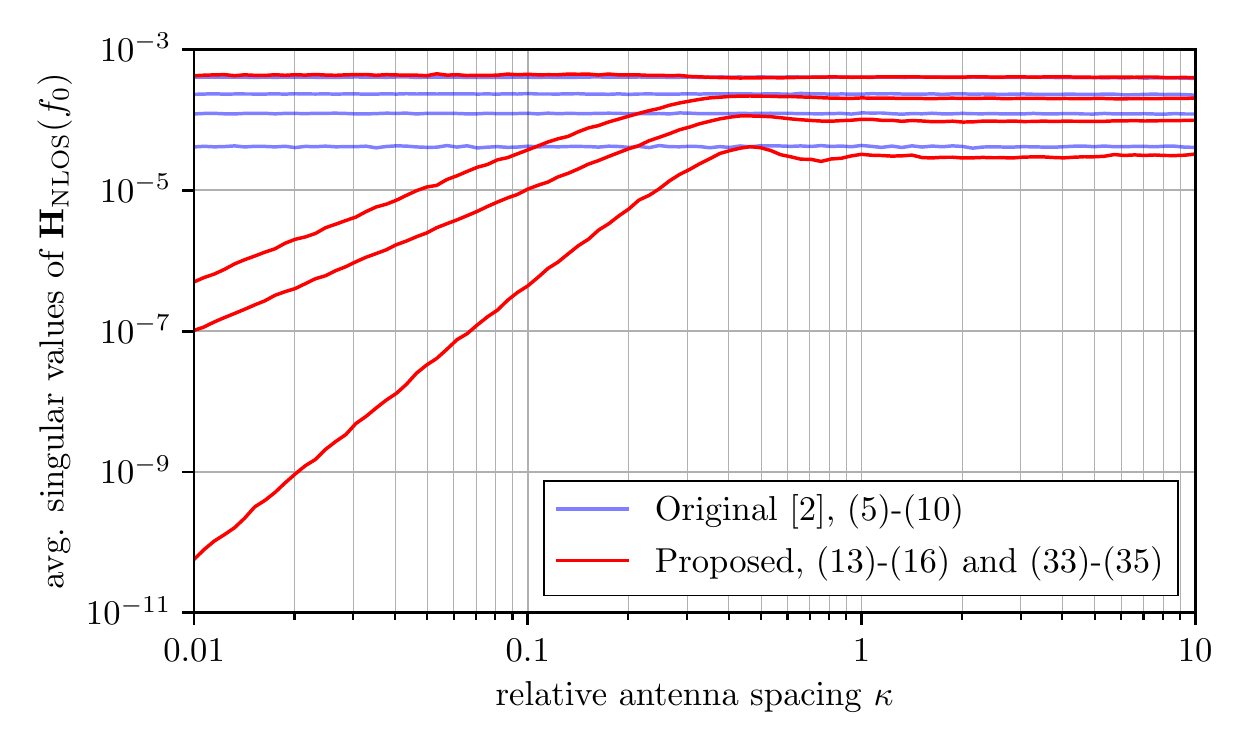}
      \label{fig:sv_ka}
  }%
  \hfil%
  \subfloat[]{
      \includegraphics[width=0.95\columnwidth]{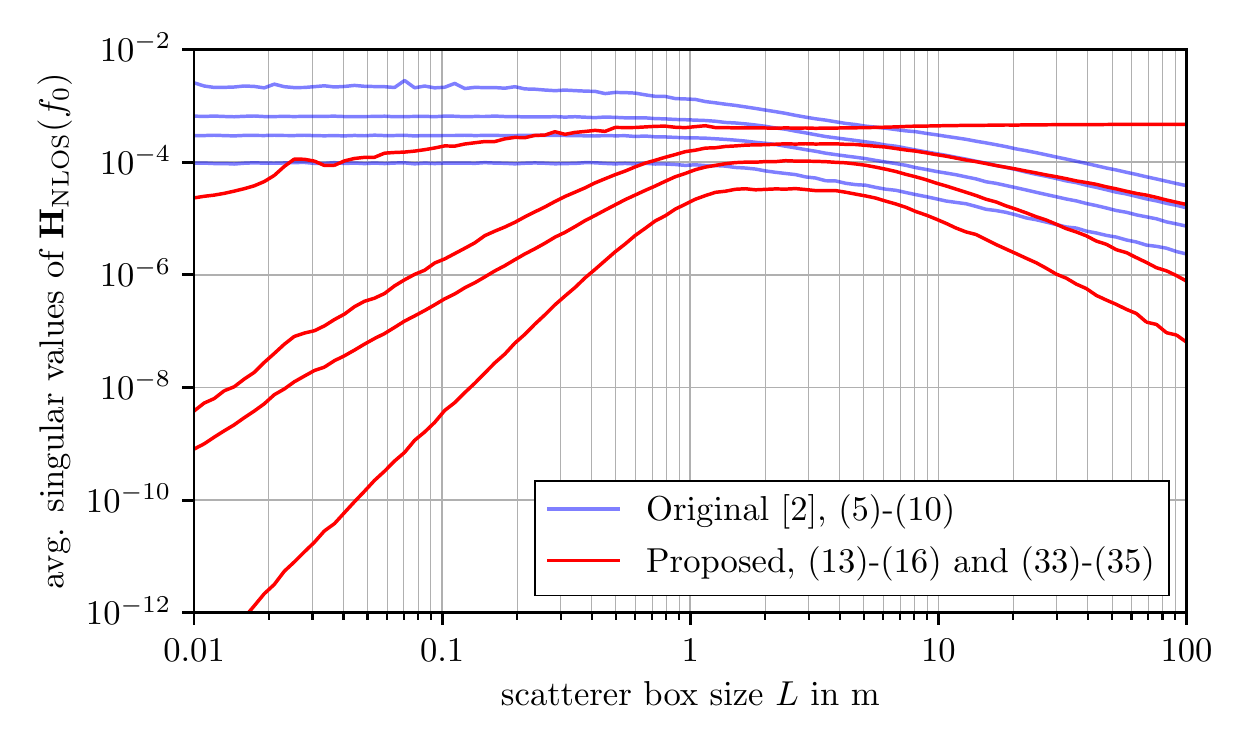}
      \label{fig:sv_l}
  }
  \caption{%
    Comparison of the average singular values of $4\times4$ MIMO realizations of the NLOS channel between the original \refParaOld{} and new \refParaNew{} parametrizations using $M=1000$ different realizations.
    (a) Singular values over the antenna spacing factor $\kappa$. Neighboring antennas within the Tx and Rx arrays are spaced $\kappa c_0 / f_0$ apart. For very small $\kappa$ the channel effectively becomes a SISO channel, and thus only one singular value should remain.
    (b) Singular values over the scatterer box size $L$. The scatterers are distributed in a cube with side length $L$ centered in the middle of Tx and Rx.
  }\label{fig:sv}
\end{figure*}

We are now going to derive an approximate expression for the $K$-factor in \cref{eq:k} for the new parametrization.
Starting from \cref{eq:k}, the LOS power $P_{\mathrm{LOS}}$ is straight forward to calculate.
Inserting \cref{eq:new-d} into \cref{eq:hlos} and \cref{eq:k} after integration yields
\begin{equation}\label{eq:power-los}
  P_{\mathrm{LOS}} = \frac{\varepsilon_D (\fmax - \fmin)}{{(4 \pi)}^2 \fmax \fmin} \sum_{m = 1}^\Nr \sum_{n = 1}^\Nt \frac{1}{{\tau_{D,mn}^2}}.
\end{equation}

The calculation of $P_{\mathrm{NLOS}}$ is more involved.
In addition to the assumptions from above, we require that the positions of the scatterers and the phase terms $\exp(-j 2 \pi \tau_{A,mn} f)$ in \cref{eq:new-t,eq:new-r,eq:new-b} are i.i.d..
However, the delays $\tau_{A,mn}$ depend the positions of the scatterers and antennas and are thus not necessarily independent.
Therefore, we have to be satisfied with approximate statistical independence between the phase terms $\exp(-j 2 \pi \tau_{A,mn} f)$, which holds if $\exp\left(-j 2 \pi \left(\tau_{A,mn} - \Exp{\tau_{A,mn}}\right) f\right)$ is uniformly distributed on the unit circle.
This in turn is likely if
\begin{equation}\label{eq:assumption-taf}
  \sqrt{\Var{\tau_A}} \fmin \gg 1, \qquad \forall A \in \{T, R, B\},
\end{equation}
where $\Var{\tau_A}$ is the variance of the delays of the corresponding matrix and $\fmin$ is the lowest frequency of interest.
In the interest of a simpler notation, we omit the explicit dependency on $f$ wherever applicable for the rest of this subsection.
As further notational aide we introduce $\m{S} = \sum_{k = 0}^{\infty} {\m{B}}^k$.
The first goal is to get an approximate expression for $\mathrm{E}\big\{\norm{\Hnlos}_\mathrm{F}^2\big\}$.
Starting with a single element, we obtain
\begin{multline}\label{eq:exphnlos}
  \Exp{{|H_{\mathrm{NLOS},mn}|}^2} = \Exp{{\left|\sum_{i = 1}^\Ns \sum_{j = 1}^\Ns R_{mi} S_{ij} T_{jn}\right|}^2} \\
  = \sum_{i = 1}^\Ns \sum_{j = 1}^\Ns \Exp{{|R_{mi}|}^2 {|S_{ij}|}^2 {|T_{jn}|}^2} = P_{H,1} + P_{H,2},
\end{multline}
where we used the assumption that the phases of $\m{R}$ and $\m{T}$, as well as the scatterer positions, are statistically independent.
We define
\begin{equation}\label{eq:ph1}
  P_{H,1} = \sum_{i = 1}^\Ns \Exp{{|S_{ii}|}^2} \Exp{{|R_{mi}|}^2 {|T_{in}|}^2},
\end{equation}
which relates to the diagonal elements of $\m{S}$ and
\begin{equation}\label{eq:ph2}
  P_{H,2} = \sum_{\substack{i,j = 1\\i \neq j}}^\Ns \Exp{{|R_{mi}|}^2} \Exp{{|S_{ij}|}^2} \Exp{{|T_{jn}|}^2},
\end{equation}
which relates to the non-diagonal elements of $\m{S}$.
The split of the double sum in \cref{eq:exphnlos} into \cref{eq:ph1} and \cref{eq:ph2} is necessary because the statistics of $\Hnlos$ are different for diagonal and non-diagonal elements.
Using the i.i.d. assumptions regarding scatterer positions and that the antennas within the arrays are relatively close together, the delays between antennas and scatterers are approximately the same across all combinations of scatterers and antennas.
Thus,
\begin{align}
  \Exp{{|T_{mn}|}^2} &\approx \Exp{{\left(\sqrt{\frac{\alpha}{f}} e^{\tau_T \gamma}\right)}^2} = \frac{\alpha}{f} M_{\tau_T}(2 \gamma), \label{eq:exp-t} \\
  \Exp{{|R_{mn}|}^2} &\approx \Exp{{\left(\sqrt{\frac{\alpha}{f}} e^{\tau_R \gamma}\right)}^2} = \frac{\alpha}{f} M_{\tau_R}(2 \gamma), \label{eq:exp-r}
\end{align}
where $\tau_T$ and $\tau_R$ are the random delays from Tx to scatterers and scatterers to Rx respectively.
We use $M_X(t) = \Exp{e^{Xt}}$ to denote the moment generating function.
Similarly, we obtain
\begin{equation}\label{eq:exp-tr}
  \Exp{{|T_{mn}|}^2 {|R_{mn}|}^2} \approx {\left(\frac{\alpha}{f}\right)}^2 M_{\tau_T + \tau_R}(2 \gamma)
\end{equation}
where $\tau_T + \tau_R$ is the joint Tx to scatterer to Rx delay.
It can be shown that the diagonal and non-diagonal elements of $\m{B}^k$ have different statistics, i.e.,
\begin{equation}
  \Exp{\big|\big(\m{B}^k\big)_{mn}\big|^2} =
  \begin{cases}
    P_{B,k,1}, & m = n \\
    P_{B,k,2}, & m \neq n
  \end{cases}.
\end{equation}
Furthermore, assuming \cref{eq:assumption-taf}, it approximately holds that
\begin{equation}\label{eq:recursion-pb}
  \begin{bmatrix}
    P_{B,k+1,1} \\ P_{B,k+1,2}
  \end{bmatrix}
  =
  \begin{bmatrix}
    0 & (\Ns - 1) \beta^2 \\
    \beta^2 & (\Ns -2) \beta^2
  \end{bmatrix}
  \begin{bmatrix}
    P_{B,k,1} \\ P_{B,k,2}
  \end{bmatrix}.
\end{equation}
In a similar fashion it can be shown that the phase terms of the elements in $\m{B}^k$ and $\m{B}^{k+1}$ are approximately statistically independent.
Together with $\m{S} = \sum_{k = 0}^{\infty} \m{B}^k$ we thus find
\begin{equation}\label{eq:exp-s}
  \Exp{{|S_{mn}|}^2} =
  \begin{cases}
    P_{S,1} = \sum_{k = 0}^\infty P_{B,k,1}, & m = n \\
    P_{S,2} = \sum_{k = 0}^\infty P_{B,k,2}, & m \neq n
  \end{cases}.
\end{equation}
By solving the Neumann series in \cref{eq:exp-s} involving the matrix from \cref{eq:recursion-pb} with $P_{B,0,1} = 1$ and $P_{B,0,2} = 0$, we obtain
\begin{equation}\label{eq:ps1}
  P_{S,1} = \left(1 - (\Ns - 1) \frac{\beta^2}{1 + \beta^2} \right){\left(1 - (\Ns - 1) \beta^2\right)}^{-1},
\end{equation}
\begin{equation}\label{eq:ps2}
  P_{S,2} = \frac{\beta^2}{1 + \beta^2}{\left(1 - (\Ns - 1) \beta^2\right)}^{-1}.
\end{equation}
We now combine \cref{eq:exphnlos}--\cref{eq:ps2}, insert back into \cref{eq:k}, solve the integration and finally find
\begin{equation}\label{eq:power-nlos}
  P_{\mathrm{NLOS}} \approx \frac{\alpha^2 \Nr \Nt \Ns (\fmax - \fmin)}{\fmin\fmax} Q(\beta, \gamma)
\end{equation}
together with
\begin{multline}\label{eq:q}
  Q(\beta, \gamma) = \frac{1}{1 - (\Ns - 1) \beta^2} \bigg( M_{\tau_R + \tau_T}(2 \gamma) + \\
  \frac{(\Ns - 1) \beta^2}{1 + \beta^2} \Big( M_{\tau_R}(2 \gamma) M_{\tau_T}(2 \gamma) - M_{\tau_R + \tau_T}(2 \gamma) \Big) \bigg).
\end{multline}

% -----------------------------------------------------------------------------------------------------
\subsection{Computing the Parameters}

What remains is to find expressions for $\alpha$, $\beta$ and $\gamma$, based on $\rho_1$, $\rho_2$ and $K$.
Starting with $\gamma$, we convert $\rho_2$ in \si{dB\per{}s} from \cref{eq:sv} to the form required by \cref{eq:new-t,eq:new-r} and find
\begin{equation}\label{eq:gamma}
  \gamma = \frac{\rho_2}{10 \log e}.
\end{equation}
Every scattering bounce corresponds to one cluster in the impulse response.
On average every such bounce delays the signal by $\Exp{\tau_B}$ and attenuates its power by $(\Ns - 1) \beta^2$, where $\tau_B$ is the random delay between the scatterers.
Thus,
\begin{equation}\label{eq:beta}
  \beta \approx \sqrt{\frac{1}{\Ns - 1}} 10^{\frac{\Exp{\tau_B} \rho_1}{10}},
\end{equation}
were, $\rho_1$ is the cluster decay rate in \si{dB\per{}s}.
Finally, inserting \cref{eq:power-los} and \cref{eq:power-nlos} into \cref{eq:k} and solving for alpha yields
\begin{equation}\label{eq:alpha}
  \alpha \approx \sqrt{\frac{\varepsilon_D \sum_{m = 1}^\Nr \sum_{n = 1}^\Nt \tau_{D,mn}^{-2}}{{(4 \pi)}^2 K \Nr \Nt \Ns Q(\beta, \gamma)}},
\end{equation}
with $Q(\beta, \gamma)$ from \cref{eq:q}.

% -----------------------------------------------------------------------------------------------------
\section{Simulation Results}\label{sec:sim}
% -----------------------------------------------------------------------------------------------------
In this section we provide simulation results comparing the original parametrization \refParaOld{} to our newly proposed parametrization \refParaNew{} and \cref{eq:gamma,eq:beta,eq:alpha}.
We used a $4 \times 4$ MIMO setup with Tx and Rx being $2 \times 2$ arrays parallel to each other at a distance of $D_0$.
The antennas are omnidirectional, and neighboring antennas within the arrays are spaced $\kappa_0 c_0 / f_0$ apart.
The $\Ns$ scatterers are uniformly distributed in a cube with side length $L$ centered between Tx and Rx.
To avoid the original parametrization's stability issues, we imposed a minimum distance between the scatterers themselves and scatterers and antennas.
We adjusted the cluster decay rates $\rho_1$ of both models to be the same.
The $K$-factor of the original model was found numerically, and the new parametrization was set accordingly.
The values for $M_{\tau_R}(2 \gamma)$, $M_{\tau_T}(2 \gamma)$ and $M_{\tau_R + \tau_T}(2 \gamma)$ were estimated based on the actual realizations.

\begin{table}
  \setlength{\extrarowheight}{2pt}
  \caption{Simulation parameters}
  \label{tab:parameters}
  \centering
  \footnotesize
  \begin{tabular}{|lll|}
    \hline
    Parameter & Symbol & Value \\
    \hline
    Tx and Rx & & 2 by 2 quadratic arrays\\
    Antennas & & omnidirectional \\
    Antenna spacing & & $\kappa_0 c_0 / f_0$ \\
    Antenna spacing factor & $\kappa_0$ & 1 \\
    Frequency & $f_0$ & \SI{5}{GHz} \\
    Tx -- Rx distance & $D_0$ & \SI{3}{m} \\
    Number of scatterers & $\Ns$ & 10 \\
    Scatterer box size & $L_0$ & \SI{5}{m} \\
    Minimum scatterer distance & & \SI{1.5}{m} \\
    Link indicators & $\varepsilon_{A,mn}$ & 1 \\
    Cluster decay rate & $\rho_1$ & \SI{-1}{dB\per{}ns} \\
    Ray decay rate & $\rho_2$ & \SI{-2}{dB\per{}ns} \\
    $K$-factor & $K$ & 180 \\
    Realizations & $M$ & 1000 \\
    \hline
  \end{tabular}
\end{table}

First, we give a qualitative comparison between realizations of the models shown in \cref{fig:realization}.
The impulse responses are approximately computed by multiplying the channel transfer matrix with a Hann window and subsequent Fourier transformation to the time domain, as was done in \cite{pedersen12}.
We observe that both models seem to have comparable statistics in the frequency domain and a comparable cluster decay rate in the time domain.
Also of note is the visible exponential decay of the new parametrization within a cluster with a rate of $\rho_2$.

Next, \cref{fig:sv} shows the average singular values of the MIMO channels.
In \cref{fig:sv_ka} we vary the antenna spacing factor $\kappa$ (instead of $\kappa_0$).
The difference between the two parametrizations stems from the reduced number of additional random phases in the newly proposed parametrization.
The channel from the new parametrization loses DoF, i.e., some singular values become notably smaller as the antennas within the arrays get closer to each other and thus become more correlated.
Intuitively, a MIMO array loses DoF as the antennas get closer than $\lambda / 2$.
However, the original parametrization fails to show this behavior.
Similarly, in \cref{fig:sv_l}, the side length of the cube containing the scatterers $L$ (instead of $L_0$) is varied.
Here, the minimum required distance between the scatterers is set to zero.
The new parametrization can replicate the expected loss of DoF of the transfer function as $L$ becomes small, i.e., the scatterers get closer to each other, and thus the NLOS channel effectively becomes a keyhole.
On the other hand, we can also observe potentially unwanted, different behavior of the two parametrizations when $L$ becomes large.
This stems from the difference in how delays are weighted in $\m{T}(f)$ and $\m{R}(f)$, i.e., polynomial versus exponential.

\begin{figure}
  \centering
  \includegraphics[width=0.95\columnwidth]{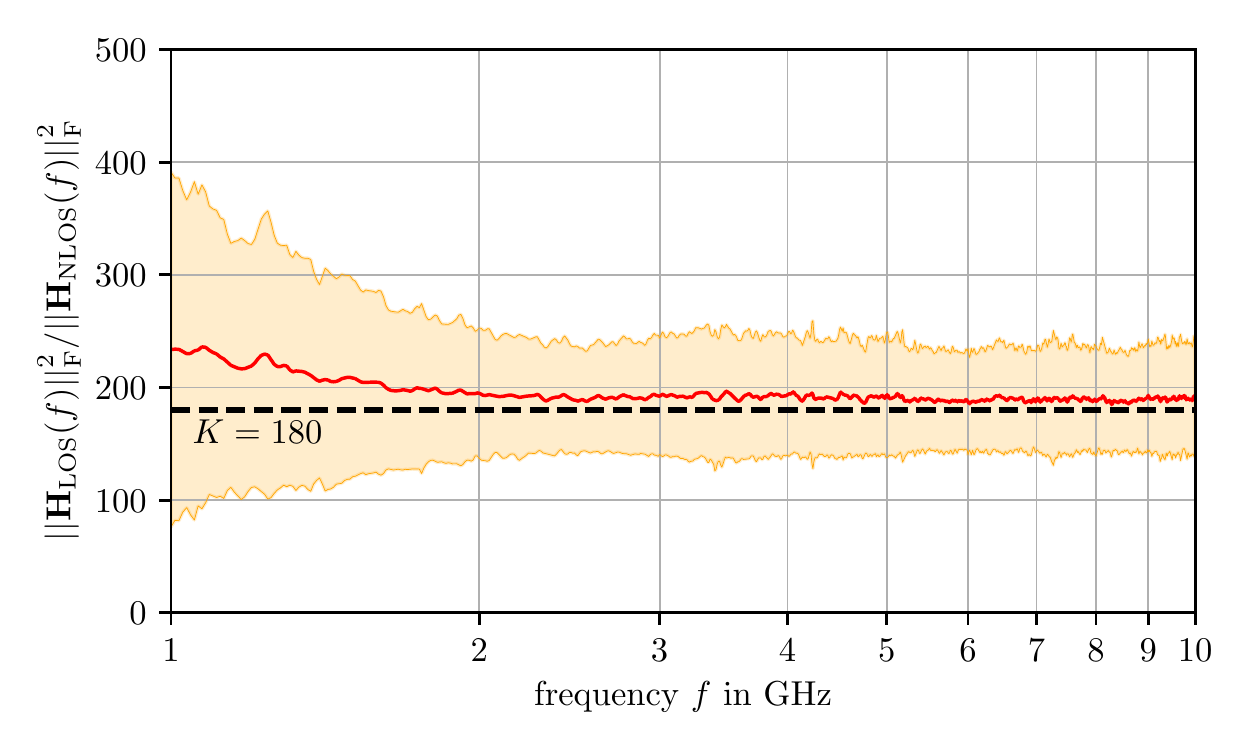} 
  \caption{%
    Fraction of the norms of the LOS and NLOS transfer matrices based on the new parametrization \refParaNew{} together with \cref{eq:gamma,eq:beta,eq:alpha}.
    The center line is the mean based on $M=1000$ simulation runs, the shaded area indicates the $\pm\sigma$ interval.
    The dashed line shows the target $K=180$.
  }
  \label{fig:k_factor}
\end{figure}

Finally, in \cref{fig:k_factor} the fraction $\norm{\m{H}_{\mathrm{LOS}}(f)}_\mathrm{F}^2 / \norm{\m{H}_{\mathrm{NLOS}}(f)}_\mathrm{F}^2$ over the frequency compared to the target $K$ is shown for the new parametrization.
While for higher frequencies the simulations are in acceptable aggreement with the target, we see a notable deterioration for lower frequencies.
However, this is expected as our derivations in \cref{ssec:deriv-k} assumed high enough frequencies such that the resulting phases are approximately statistically independent.
Across all our simulations we got $\max\big(\sqrt{\Var{\tau}}\big) \approx \SI{4}{ns}$, for any random delay $\tau$, and according to \cref{fig:k_factor} the approximation starts to perform badly for $f < \SI{2}{GHz}$.
Thus, based on our simulations it seems that
\begin{equation}
  \sqrt{\Var{\tau}} \fmin > 8
\end{equation}
is an acceptable condition for the validity of \cref{eq:beta} and \cref{eq:alpha}.

% -----------------------------------------------------------------------------------------------------
\section{Conclusion}
% -----------------------------------------------------------------------------------------------------
We presented a new parametrization for PGs that orients itself on the SV model's doubly exponential decay.
We showed how to compute the PG parameters from the easier to use $K$-factor and decay rates and verified our approximations by simulations.
Finally, we showed that a MIMO channel based on our new parametrization loses DoF as the channel becomes spatially correlated, which is the intuitive behavior.
In particular, the channel loses DoF if the antennas within the arrays get close or the scatterers are distributed in a small volume.

% -----------------------------------------------------------------------------------------------------
\section*{Acknowledgment}
% -----------------------------------------------------------------------------------------------------
The financial support by the Austrian Federal Ministry for Digital and Economic Affairs and the National Foundation for Research, Technology and Development is gratefully acknowledged.

% -----------------------------------------------------------------------------------------------------
% Restore old commands
\let\v\oldv